\documentclass{cimento}

\title{Quantum mechanics at high school: an online laboratory on wave-particle duality}
\author{M.~Tuveri\from{ins:x}\from{ins:y}\thanks{Corresponding author, e-mail: matteo.tuveri[at]ca.infn.it.}\ETC,
D.~Fadda\from{ins:w},
\atque
C.M.~Carbonaro\from{ins:x}}
\instlist{\inst{ins:x} Dipartimento di Fisica, Universit\`a di Cagliari - Cagliari, Italy
  \inst{ins:y} INFN, Sezione di Cagliari - Cagliari, Italy
    \inst{ins:w} Dipartimento di Psicologia, Pedagogia e Filosofia, Universit\`a di Cagliari - Cagliari, Italy}
\begin{document}

\maketitle

\begin{abstract}
The interest in studying quantum mechanics is always increasing in our society and schools. Especially in the latter case, this leads researchers to implement suitable actions to meet social needs of knowledge of quantum physics. We present an online laboratory on wave-particle duality for high school students (17-19 years old). The activity has been carried out in the period December 2021 - May 2022 at the Physics Department of the University of Cagliari and more than 100 students from different high schools in Sardinia have been involved. We will show the design of the activity and the experiments performed. We will show and discuss qualitatively results about a satisfaction questionnaire. A brief discussion about motivational issues will be done. 
\end{abstract}

\section{Introduction}
Quantum mechanics is around us, and the interest in studying this subject is increasing in our society. For example, topics related to quantum physics are now part of high schools’ programs. Newspapers, tv shows and science communication profiles on social media often talk about quantum technologies around us. To learn and to be informed about quantum mechanics and its application in our research as well as in our everyday life is important for cultural reasons and to become consciousness citizens~\cite{ref:Michelini2021}. Fron this point of view, researchers play an important role in society, they have to implement suitable actions to meet social needs of knowledge of quantum physics. 

Starting from schools, many strategies can be used to face with the quantum world, focusing on technological aspects~\cite{ref:Oss2015, ref:Malgieri2019} or on historical and informal ones~\cite{ref:Grimellini2004}. The educational content of these approaches can focus on different subjects, from conceptual and linguistic aspects~\cite{ref:Levrini2008, ref:Singh2001}, where the language, both the natural and the mathematical one, is used as an instrument to introduce the peculiar features of the quantum world~\cite{ref:Singh2006}. 
Another possibility can be to focus on one of the main conceptual issues of quantum mechanics, that is the wave-particle duality and develop suitable learning strategies to highlight the manifestation of the dual nature of matter and light. %How to implement such strategies in an online environment?

In this paper we present a laboratory on wave-particle duality for high school students (17-19 years old). The experimental activities had been carried out in the period December 2021 - May 2022 at the Physics Department of the University of Cagliari. More than one hundred students from different high schools in Sardinia have been involved, whose participation was online due to the pandemic. Main aim is to show the design of the activity and the experiments performed.
Inspired by previous research on this field~\cite{ref:Michelini2021}, we also wrote a research questionnaire to understand how the online laboratory affected students' motivation and interest in physics, their vision of the scientific method and the influence of the laboratory on their understanding of physics and the concepts studied at school. A detailed analysis will appear in a forthcoming paper. Research methodology and a qualitative analysis of data are presented.

\section{Methods}
The main topic of the laboratory was the wave-particle duality. We focus on waves (mechanical and electromagnetic) 
and their properties, as well as on particular macroscopic properties and phenomenology of matter (such as scattering). 
Four different experiments dealing with the undulatory and particle properties of matter were shown and discussed. 

The first experiment deal with mechanical waves propagating in a fluid. 
In this case, researchers focused on diffraction as a key phenomenon to 
introduce the dual behavior of matter according to the experimental set-up and conditions. 
%Indeed, the concept of slit and of its dimensions with respect to the wavelength of the wave is introduced.   
The second experiment was a flipper-like apparatus, with marbles hitting a screen passing through a slit. 
This was to explain and show the particle behavior of matter, that is that massive particles and, in general, 
macroscopic objects (with a length of the order of centimeters or more) do not diffract. 
Also in this case, researchers focused on the phenomenon of elastic scattering 
and the relationship between the dimension of the marbles and the slit. 
The third experiment concerned the diffraction of light. 
A red light emitted by a laser (with a wavelength of about 650 nm) passes through some lenses and a slit
to be coherently collimated in a beam. 
The slit can be opened or closed until its size becomes comparable with the laser wavelength. 
Then, diffraction occurs. 
Finally, in the fourth experiment, researchers showed electron diffraction through a suitable 
experimental set-up (the electron diffraction system build by Phywe). %~\footnote{See \url{https://www.phywe.com/experiments-sets/nobel-prize-experiments/electron-diffraction_9532_10463/} for details.}.
In this case, the diffraction manifests with rings on a fluorescent screen. 
This experiment shows that, 
under suitable conditions, that is an electron passing through 
a slit (graphite planes) of dimensions comparable with its wavelength, even what is 
typically thought as a particle manifest an undulatory phenomenology. 
To also show that in this process, the electron does not loose its charge, we used a magnet 
to move the diffraction figure along all the screen. 

The methodological structure of the laboratory was as follows. 
Firstly, an introductory game was proposed using the "Quizziz" platform to qualitatively measure 
students' expectations about phenomena showed during the activity. The questions had not any evaluation 
intent, rather they just measured their feeling or previous knowledge (especially for students 
attending the last years in high school) on the subject. 
%Questions spanned from their expectations 
%about what happened when some mechanical wave, marbles, the light of a laser and the electrons 
%in the experiments previously cited passes through a slit of suitable dimensions. 
This activity lasts ten minutes. 
After that, the experimental activities started (duration: 40 minutes). 
%During the presentation of the four experiments, some historical reference to debates and discoveries 
%related to the corpuscular and undulatory nature of light, as well as, on the Einstein's work on photons was given. 
%Moreover, researchers also mentioned some technological and everyday life application of such phenomena. 
%For example, this is the case of the photoelectric effect in the automatic doors opening in supermarket, or the 
%wi-fi or radio communication of smartphone (thanks to the diffraction of electromagnetic waves). 
%This was made to contextualize the phenomena students were observing to our everyday life., 
%giving them the sense of cultural revolution of the discoveries made in the first years of the nineteenth century. 
The laboratory ended with a general recap on the physics dealt and the results of the introductory game 
was discussed in the light of phenomena observed. 
%where physics is taught according a procedure very similar to the scientific method: phenomenon, 
%expectations on it (hypotheses), observation of a phenomenology (to verify or falsify the hypotheses) 
%and, only at the end, the model to describe (formally or not formally) the phenomenon. 
We left a detailed discussion on the pedagogical approach to a future paper, where further details will be given. 

The total duration of the activity was about one hour. Contents where targeted: the more the participants were 
young the less technicalities and details were inserted in the discussion. Despite the online environment, a certain level of 
interaction with the class through the mediation of the teacher was guaranteed. 

Participants to the synchronous online session were 104 high-school students attending 
the last three years of Lyceums in Sardinia (in the metropolitan area of Cagliari, 1 "humanities" and 5 "scientific"). 
The total number of participants was obtained by summing the in-class counting made by teachers 
once the synchronous session started. Teachers and students attended the online laboratory from their classrooms and 
a Zoom connection was established, with cameras filming the researchers and the experiments 
connected to a laptop with a Raspberry system. Every class attended the laboratory separately, 
thus the total number of meeting was 6. 

We wrote a satisfaction questionnaire to investigate students' feedback on their 
experience with the online laboratory (2 items); on the influence of the laboratory on their understanding of 
physics and the concepts studied at school (2 items); on their vision of science and of the scientific method (3 items); 
on the interaction with researchers (2 items). 
Students could answer by using a 5-point Likert scale, from 1 (completely disagree) to 5 (completely agree).
For each class, data were collected from 15 days to one month after the end of the synchronous meeting. 
The questionnaire was written in Italian and imported in Microsoft Forms. The teacher distributed it as a link via
email to students. Students’ participation was voluntary with no positive or negative inducements. 
The questionnaire was anonymous, no information on gender or class was obtained. 
The number of answers collected was 104.

%Data were qualitatively analyzed by calculating means and standard deviation. A factor analysis was also performed. 
In the following, we just show and discuss the qualitative results related to students's satisfaction
on the 4 domains cited above.  

\section{Results}
Concerning students' feedback on their experience with the online laboratory, most of them (78.8\%) 
affirmed that the topics of the lab were interesting. One half of them (51.1\%) thought the lab fostered their 
curiosity on the topics of the lab, whereas one third of the sample (30.8\%) stay neutral on this item. 
Concerning the influence of the laboratory on their understanding of 
physics and the concepts studied at school, 46.2\% of students affirmed that thanks to the lab, 
he/she could explore the physics phenomena he/she was studying at school. 
The 29.8\% of the sample was neutral. 
The capacity of the lab to engage students in in studying physics was rated as good by the 66.3\% of students. 

Concerning their vision of science and of the scientific method, 
the 68.2\% of students affirmed that the lab helps them in understanding the importance of taking, analyzing, collect, and interpret the data. 
Most of them (67.3\%) affirmed that the lab helps them in understanding how to carry a scientific research on. The same happens 
when we asked students if the lab allowed them to think about and explain the observed phenomena: 
in this case, the 57.5\% of students agreed with this item. 
Concerning the interaction with researchers, the majority of them (78.8\%) affirmed that the interactions with researchers were useful. 
Finally, the item: "attending the remote lab with the researcher helps me in understanding the experiment" was positively rated by the 65.3\% 
of the sample.

\section{Discussion and conclusions}
The qualitative results on students' interest and curiosity towards physics, 
as well as on their motivation in participating to the online activity are encouraging.   
Moreover students appreciated to interact with researchers even in an online environment. Most of them also affirmed  
that the interaction with researchers was also helpful in understanding the physics behind the experiments. 
This result suggests that interaction is a key point in learning and in outreaching activities, too. %~\cite{}. 
Another interesting result is that students affirmed that our initiative helps them in 
understanding the scientific method. The laboratory seemed to have a discrete influence also 
on students' understanding of physics concepts studied at school. 

Some criticality emerged: teacher-mediated interaction between researchers and students 
did not encourage a constant and active participation of the latter to the lecture. 
Students appeared to be scared by a possible judgements of their teacher if they were wrong in talking with researchers.
This is a crucial point to be faced up in order to find strategies to implement online learning of 
physics in school in curricular timetable. A possible solution can be to include teachers 
in the design of the activity, thus making them co-authors of the project. In this 
sense, if they already know what researchers will show, then, they can explain concepts to their 
class during the synchronous activity, e.g., when internet connection arises or when he/she think 
this is needed.
% From the motivational point of view, the presence of young researchers or university 
%students would increase interaction, making the activity more appealing for students. 
%More experiments and measures are needed to involve students in a proper "research based activity", 
%where students can guide the researchers' hands in quantifying the phenomena they are seeing through 
%a display. For example, this can be done in the case of electron diffraction, where the measure of the 
%Planck constant can be performed.  

For the future, we hope to increase the sample to better the statistics and to understand the efficacy 
of our methodology, possibly introducing a quantitative measure of students' learning of concepts 
proposed in the laboratory with a pre and a post questionnaire. 
We are also planning to implement the laboratory in a proper education 
and learning platform allowing us to study all the steps of participants' online learning.
This platform will allow us to follow the students also in the asynchronous phase, when they re-elaborate 
the supplementary and learning material uploaded on the platform and 
focus on the content of the laboratory. All these activities are left for a future study.

\acknowledgments
The author acknowledge faculties, teachers and students who participated in the study.

\end{document}